\UseRawInputEncoding
\documentclass[aps,prb,groupedaddress,twocolumn,showpacs]{revtex4-1}
\usepackage[dvips]{graphicx}
\usepackage{dcolumn}
\usepackage{bm}
\usepackage{amssymb}
\usepackage{epstopdf}
\usepackage{amsmath}
\usepackage{color}
\usepackage[dvipdfmx]{hyperref}
\hypersetup{
 setpagesize = false,
 colorlinks = true,
 citecolor = blue,
 linkcolor = blue,
 urlcolor = blue,
 }
\usepackage{CJK}
\begin{document}

\title{High-temperature ferromagnetism and strong $\pi$-conjugation feature in two-dimensional manganese tetranitride}
\author{Ming Yan$^{1}$}
\author{Z. Y. Xie$^{2}$}\email{Corresponding author: qingtaoxie@ruc.edu.cn}
\author{Miao Gao$^{3}$}\email{Corresponding author: gaomiao@nbu.edu.cn}
\date{\today}
\affiliation{$^{1}$Department of Mechanical and Aerospace Engineering, Syracuse University, Syracuse, NY 13244, USA}
\affiliation{$^{2}$Department of Physics, Renmin University of China, Beijing 100872, China}
\affiliation{$^{3}$Department of Physics, School of Physical Science and Technology, Ningbo University, Zhejiang 315211, China}
\begin{abstract}
  Two-dimensional (2D) magnetic materials have attracted tremendous research interest because of the promising application in the next-generation microelectronic devices. Here, by the first-principles calculations, we propose a two-dimensional ferromagnetic material with high Curie temperature, manganese tetranitride MnN$_4$ monolayer, which is a square-planar lattice made up of only one layer of atoms.
  The structure is demonstrated to be stable by the phonon spectra and the molecular dynamic simulations, and the stability is ascribed to
  the $\pi$-d conjugation between $\pi$ orbital of N=N bond and Mn $d$ orbital.
  More interestingly, the MnN$_4$ monolayer displays robust 2D ferromagnetism, which originates from the strong exchange couplings between Mn atoms due to the $\pi$-d conjugation.
  The high critical temperature of 247 K is determined by solving the Heisenberg model with the Monte Carlo method.

\end{abstract}


\maketitle


Because of the great application prospect in the fields of electronics, information, energy, and chemical industry, two-dimensional materials have attracted extensive research interest. The quantum confinement effect due to the reduced dimension endows them exotic physical properties in comparison with their bulk counterparts.
After graphene was discovered, a variety of 2D materials have been reported in the theoretical and experimental studies \cite{Ashton2017}, including boron nitride \cite{Song2010}, silicene \cite{Lalmi2010}, transition metal dichalcogenides \cite{Coleman2011}, MXenes \cite{Naguib2011}, $etc$.
Among them, only a small fraction are magnetic compounds.
The Mermin-Wagner theorem states that long-range magnetic ordering is absent in an isotropic 2D Heisenberg system.
Based on the belief, such studies on magnetic 2D materials have been ignored or shunned, which limits the development of the magnetic 2D material field.
Until 2017, the intrinsic 2D ferromagnetism was found for the first time in CrI$_3$ \cite{Huang2017} and Cr$_2$Ge$_2$Te$_6$ \cite{Gong2017} atomic layers, in which the single-site magnetic anisotropy on Cr atoms breaks the isotropy due to the spin-orbital coupling.
Since then, the issues concerning 2D magnetism are getting more and more attention.

In terms of the thickness of 2D materials, they can be classified into two categories, one contains only one layer of atoms and another is composed of a few layers of atoms, for the examples of graphene and MoS$_2$, respectively. We call the first category \textit{single-atom-thick 2D materials}, which is related to the thickness limit of film materials and all the atoms are confined in a plane. The typical materials are the graphene, borophene, boron nitride, and $g$-C$_3$N$_4$ \cite{Lalmi2010,Mannix2015,Song2010,Groenewolt2005} while the common structural pattern is the honeycomb hexagonal lattice. The combination of delocalized $\pi$ bond and sp$^2$ orbital hybridization is the main mechanism to stabilize the planar and hexagonal geometry. And more importantly, there is the real 2D crystal field around each atom in the single-atom-thick 2D materials. It is not only distinct to the crystal field in bulk compound, but also different from that in other slab 2D materials consisting of a few layers of atoms.

Manganese is a special magnetic element because of the half-filled $d$ shell in its electronic configuration of 3$d^5$4$s^2$, 
which is one of the main magnetic ingredients for the ferromagnetic compounds, especially for the ferromagnetic 2D materials.
For instance, MnO$_2$, MnS$_2$, and MnSe$_2$ monolayers are intrinsic ferromagnetic semiconductors with the T$_c$s of 140 K, 225 K and 250 K, respectively \cite{Kan2013, Kan2014}.
Mn$_3$C$_{12}$S$_{12}$ monolayer \cite{Zhao2013} and Mn-phthalocyanine (MnPc) sheet \cite{Zhou2011} are ferromagnetic with $T_c$s of 212 K and 150 K.
Recently, hexagonal MnN and pentagonal MnN$_2$ monolayers are also be predicted to be ferromagnetic with T$_c$s of 368 K and 913 K \cite{Xu2018,Zhao2020}. We note that these T$_c$s are computed according to the Ising model, and the values are usually overestimated compared to the Heisenberg model.
Inspired by the reported manganese compounds, our question is whether there are some other 2D ferromagnetic manganese nitrides with high Curie temperature?

In this work, on the basis of the first-principles calculations and Monte Carlo simulation, we demonstrate that the single-atom-thick MnN$_4$ monolayer with a square lattice is a high-temperature ferromagneton with T$_c$ of 247 K. The $\pi$-conjugation effect play an important role both in the formation of planar geometry and in the ferromagnetic exchange coupling between two Mn atoms.

The structural and electronic properties are computed with the plane wave pseudopotential method enclosed in the VASP package, and the projector augmented-wave (PAW) pseudopotential with Perdew-Burke-Ernzerhof (PBE) exchange-correlation functional \cite{PhysRevB.47.558, PhysRevB.54.11169, PhysRevLett.77.3865, PhysRevB.50.17953} are used.
 The convergence criteria for total energy and forces are set to
be 10$^{-5}$ eV and 10$^{-3}$ eV/\AA, respectively. Plane waves with a kinetic energy cutoff of 600 eV are used to expand the valence electron wave functions.
Considering the strong correlation of Mn $d$ electrons, the GGA + U method with U$_{eff}$ = 6.3 eV is applied to compute the electronic and magnetic properties \cite{Cococcioni2005}.
The interlayer distance was 16 \AA~ and a k-point mesh was $24\times 24\times 1$ for the Brillouin zone integration.
To examine the stability, the density-functional perturbation theory (DFPT) and the supercell method in the PHONOPY program were used to calculate the phonon spectra \cite{Togo2015}.
In the ab initio molecular dynamics simulations,
the 3 $\times$ 3 $\times$ 1 supercells were employed and the temperature was kept at 1000 K for 5 ps with a time step of 1 fs in the canonical ensemble (NVT) \cite{Martyna1992}.


\begin{figure}
\begin{center}
\includegraphics[width=7.5cm]{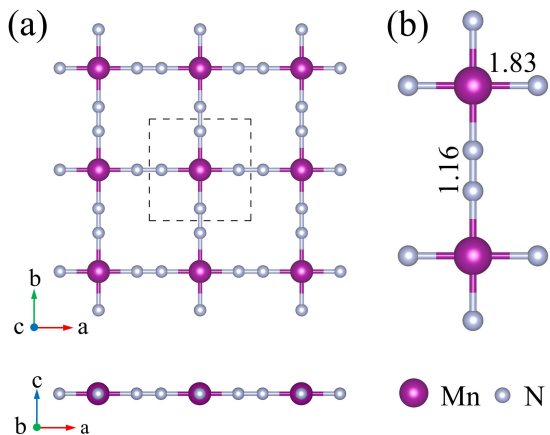}
\caption{(a) Top view and side view of square-MnN$_4$ monolayer. The unit cell is marked by a dashed line square. (b) The bond lengthes of N=N and Mn-N bonds are labeled.
 } \label{struct}
\end{center}
\end{figure}

The top view and side view of MnN$_4$ structure is displayed in Fig.\ref{struct}(a).
The structure has a square lattice with the symmetry of plane group P4m .
The lattice parameters are $a$ = $b$ = 4.83 \AA~. The lengths of Mn-N bond and N=N bond are 1.83 \AA~ and 1.16 \AA, shown in Fig.\ref{struct}(b).
Its unit cell consists of only one MnN$_4$ moiety, as marked by a dashed line square.
These moieties are connected together by the N=N bonds to make up of a planar MnN$_4$ layer,
which is a porous monolayer with large square pores.
The Mn atom is located at the center of the square of four N atoms.


To ascertain the structural stability of the MnN$_4$ monolayer, we first compute the formation energy, and the metal manganese and N$_2$ gas are taken as the reference.  The formation energy is defined below
\begin{equation} \label{H}
E_{form} = \frac{E_{tot} - E_{Mn} - 2 E_{N_2}}{n},
\end{equation}
in which $E_{tot}$, $E_{Mn}$, and $E_{N_2}$ are the total energy, bulk manganese energy per atom, and nitrogen molecule energy, respectively.
For comparison, we also compute the formation energies of a few single-layer nitrides including hexagonal MnN, pentagonal MnN$_2$, $g$-C$_3$N$_4$, and BeN$_4$ monolayers, which have already been synthesized experimentally or investigated theoretically.
The formation energy of MnN$_4$ monolayer is 0.09 eV/atom, lower than the ones of other nitrides in Tab. \ref{Eform}, which indicates that the MnN$_4$ structure is energetically stable.

\begin{table}
	\caption{The formation energies of five 2D nitride monolayers. The unit is eV/atom.
	}
	\label{Eform}
	\renewcommand\tabcolsep{5.5pt} 
\begin{tabular*}{7.5cm}{ccccc}
       \hline
      hex-MnN & pen-MnN$_2$ & $g$-C$_3$N$_4$ & BeN$_4$  & MnN$_4$  \\
		\hline
    0.28& 0.21 & 0.35 & 0.12  & 0.09  \\
	    \hline
\end{tabular*}
\end{table}

\begin{figure}
\begin{center}
\includegraphics[width=8.50cm]{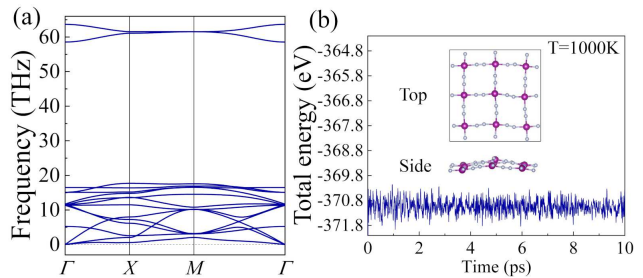}
\caption{(a) Phonon energy bands of MnN$_4$ monolayer. (b) Evolution of MnN$_4$ total energy with time at 1000 K.
 } \label{phonon-press}
\end{center}
\end{figure}

To examine the dynamic stability of MnN$_4$ monolayer, we perform the phonon calculations.
The phonon bands are shown in Fig. \ref{phonon-press}(a), and there is no imaginary frequency for their phonon modes, indicating that MnN$_4$ monolayer is dynamically stable.
 Then, we perform the first-principles molecular dynamic simulations to examine the thermal stability. The variation of total potential with respect to time is presented in Fig. \ref{phonon-press}(b), where the energy only fluctuates slightly around a constant and no distinct drop of energy emerges. The insets are the top and side views of the final structure of MnN$_4$ monolayer after 10 ps simulation at the temperature of 1000 K. It can maintain the framework and no breaking of the bonds is found, which confirm the good thermal stability of MnN$_4$ monolayer.

\begin{figure}
\begin{center}
\includegraphics[width=7.5cm]{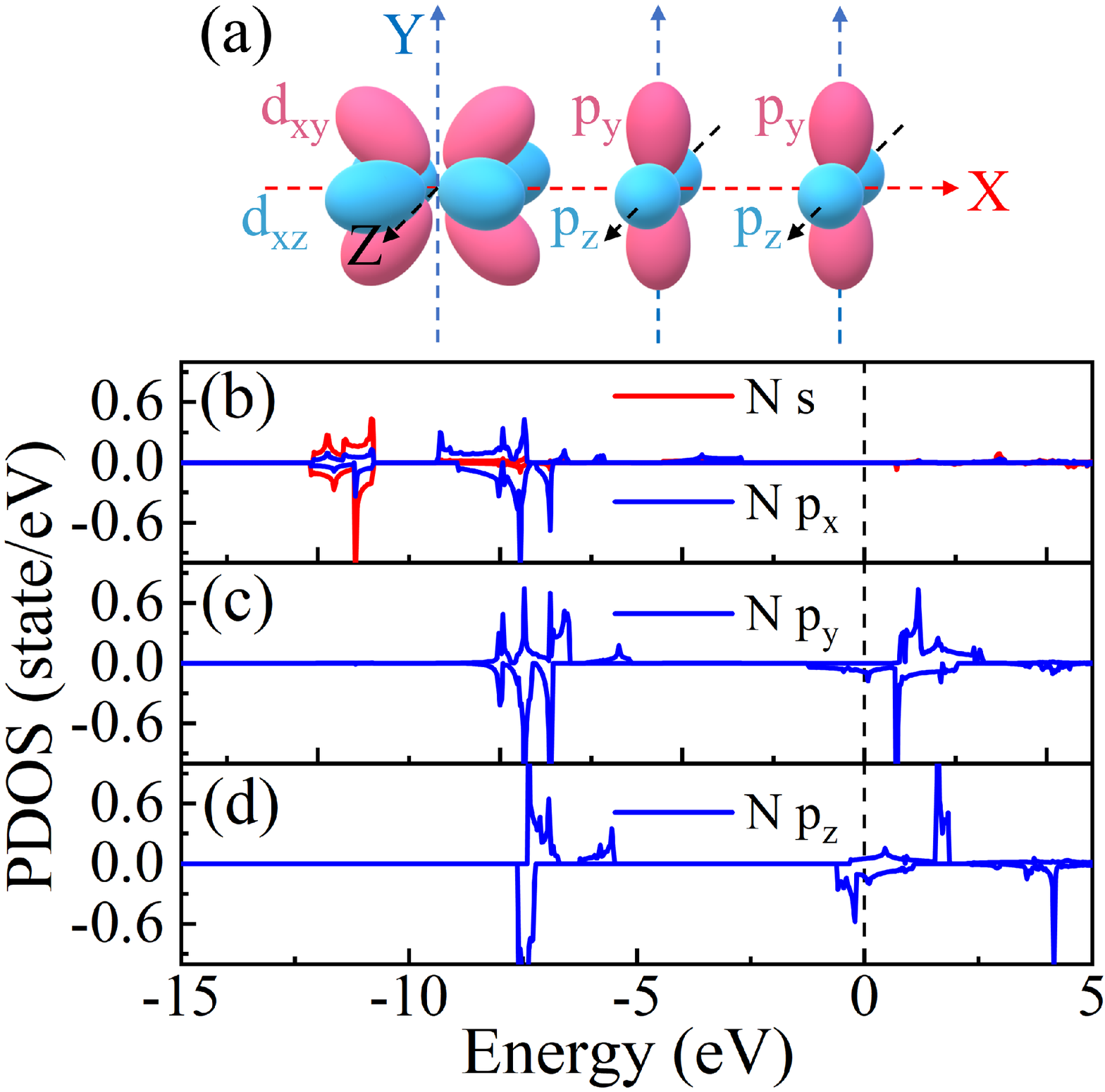}
\caption{Density of states projected on $2s$ and $2p$ orbitals of N atom. The Fermi energy is marked by the vertical dashed line.
 } \label{NNbond}
\end{center}
\begin{center}
\includegraphics[width=7.5cm]{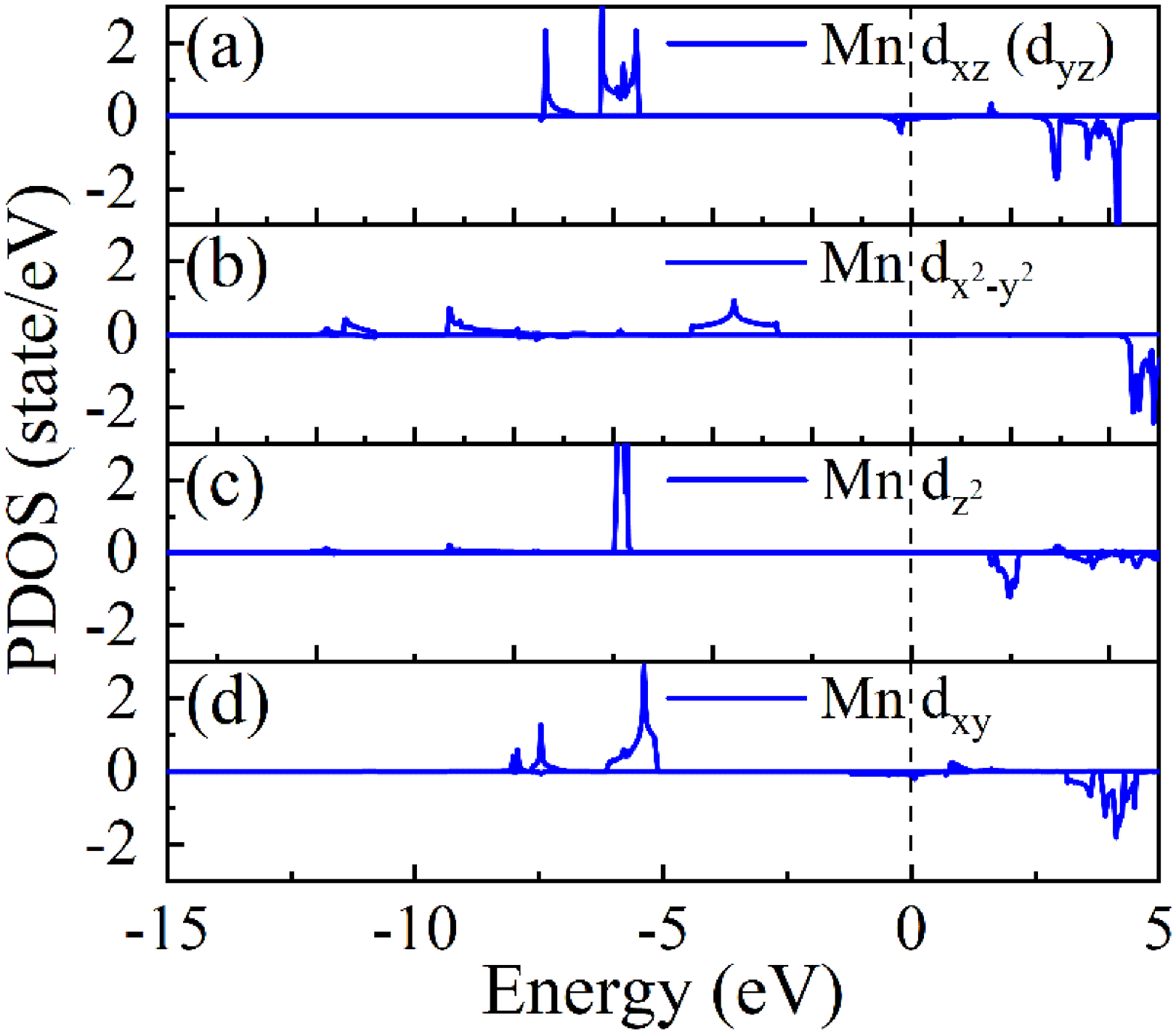}
\caption{Density of states projected on $3d$ partial orbitals of Mn atom. The Fermi energy is set to zero.
 } \label{MnPDOS}
\end{center}
\end{figure}

$\pi$ conjugation is the overlap of one $p$ orbital with another $p$ orbital (or $d$ orbital of transition metals) across an adjacent $\sigma$ bond.
We first look at the Mn-N=N-Mn chain in the MnN$_4$ monolayer.
The bonding distance of the nitrogen dimer is 1.16 \AA~ and it is close to the N=N bond length of 1.2 $\sim$ 1.3 \AA, indicating that there is a N = N double bond between two N atoms. The sketches of N $2p$ orbitals in a Mn-N=N-Mn chain along $x$ axis are exhibited in Fig. \ref{NNbond}, where two $p_y$ orbitials are aligned side by side to form a $\pi$ bond and two $p_z$ orbitials form another $\pi$ bond.
Along the Mn-N=N-Mn chain, the two $p_x$ orbital form a $\sigma$ bond between two N atoms.
Therefore, there exists strong $\pi$ conjugation originating from two $p_y$ orbitials and two $p_z$ orbitials because of the uncommon geometry of   Mn-N=N-Mn chain.

In Fig. \ref{NNbond}, the DOS of N $2s$ and $2p_x$ states are distributed in the same energy scope and there is no energetic separation between the $2s$ and $2p_x$ orbitals, which is related to the occurrence of very strong orbital hybridization, quite similar to the typical $sp^2$ hybridization of C atoms in graphene.
N $2s$ and $2p_x$ orbitals (along Mn-N-N-Mn chain) are mixed to form two hybridized orbitals, one extends to adjacent N atom to form the $\sigma$ bond and another extends to Mn atom to form the Mn-N coordination bond.
N element has five valence electrons, the three electrons occupy the $p_y$, $p_z$, and one $sp_x$ hybridized orbital, and the remaining two electrons (lone electron pair) fill the other $sp_x$ hybridized orbital.
For the Mn atom coordinated by four N atoms, the 4$s$, 4$p_x$, 4$p_y$, and 3$d_{x^2-y^2}$ orbitals is coupled each other to give rise to four $dsp^2$ hybridized orbitals. All the four orbitals lie in the $xy$ plane.
It is just the four $dsp^2$ orbital that are coupled to the $sp_x$ hybridized orbitals from the coordinated N atoms to form the coordination bonds. This is a common situation in phthalocyanine compounds and some 2D metal-organic frameworks \cite{Sauvage1982,Wang2021}.

Apart from the $\pi$ bond of N-N dimer and $dsp^2$ hybridization of Mn atom,
the $\pi$ bond from two $p_z$ orbitals is further mixed to the $d_{xz}$ orbital of Mn atom (see Fig. \ref{NNbond}), resulting in the first $\pi$-d conjugation.
On the other hand, there is a considerable overlap between the $\pi$ bond from two $p_y$ orbitals and the $d_{xy}$ orbital of Mn atom in the $xy$ plane, which leads to the second $\pi$-d conjugation.
 It is the special configuration of linear Mn-N=N-Mn chain that causes the occurrence of double $\pi$-d conjugation,
 which do not happen in the hexagonal and pentagonal networks of MnN and MnN$_2$ monolayers mentioned above.
The double $\pi$-d conjugations further elevates the strength of Mn-N bond and lowers the total energy, greatly enhances the robustness of the MnN$_4$ planar structure.

Consequently, strong $\pi$-d conjugations is not only the most distinctive feather of MnN$_4$ monolayer but also the main mechanism for the structural stability of this square-planar lattice.


Because the MnN$_4$ monolayer contains the $d$ electronic states, we utilize the GGA + U method to investigate the electronic and magnetic properties.
The Hubbard U is determined to be 6.3 eV through self-consistent calculations with linear response method \cite{Cococcioni2005}.
Fig. \ref{MnPDOS} shows the partial DOS of Mn five $3d$ orbitals.
The 3d electronic states of Mn atom have a complete spin splitting, and there is almost no overlap between the energy regions of spin-up and spin-down states. Furthermore, the spin-up states are almost entirely occupied and spin-down states are completely empty, which results in the large moment of 4.4 $\mu_B$.

\begin{figure}
\begin{center}
\includegraphics[width=7.5cm]{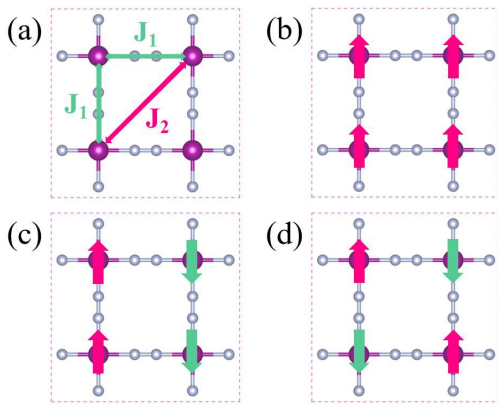}
\caption{(a) The nearest and next-nearest neighboring exchange interactions, $J_1$ and $J_2$. [(b)¨C(d)] The sketches of FM, AFM-I, and AFM-II orders.
 } \label{order}
\end{center}
\end{figure}

We build a 2 $\times$ 2 supercell to study the magnetism of the CoN$_4$ monolayer in the ground state.
Three magnetic orders, including ferromagetic order (FM) and two antiferromagnetic orders (AFM-I and AFM-II), are displayed in Fig. \ref{order}.
The energies for AFM-I and AFM-II orders are all higher than that of FM order, revealing that the ferromagnetic order is the magnetic ground state.
According to the energies, the neighboring exchange couplings can be figured out from the following expressions,
\begin{equation}
\begin{aligned}\label{J1J2}
J_1 &= \frac{1}{4}(E_{FM} - E_{AFM-II}),  \\
J_2 &= \frac{1}{8}(E_{FM} + E_{AFM-II} - 2E_{AFM-I})
\end{aligned}
\end{equation}
and the nearest and next-nearest neighboring coupling $J_1$ and $J_2$ are -24.0 meV/S$^2$ and -5.6 meV/S$^2$.
In addition, the magnetic anisotropy energy, the energy difference of Mn moment pointing to $z$ and $x$ (or $y$) axis, is computed to be 0.024 meV.

What is the mechanism of ferromagnetic exchange interaction between two Mn atoms?
Along the Mn-N=N-Mn chain, the $d_{xz}$, $p_z$, $p_z$, and $d_{xz}$ orbitals are mixed together to form the large $\pi$ orbitals.
At first, the two $d_{xz}$ are half-filled, and to make sure the electron can freely move from one $d_{xz}$ orbital to another without spin flipping, the electrons in two $d_{xz}$ orbitals must keep the same spin state.
Then, because the $d$ electron number of Mn atom is less than five, the spins are aligned in parallel according the Hund's rule.
Therefore, the Mn moments in the Mn-N=N-Mn chain are parallel, and the ferromagnetic coupling stems from the delocalized electrons associated with the $\pi$-d conjugation effect.

\begin{figure}
\begin{center}
\includegraphics[width=7.0cm]{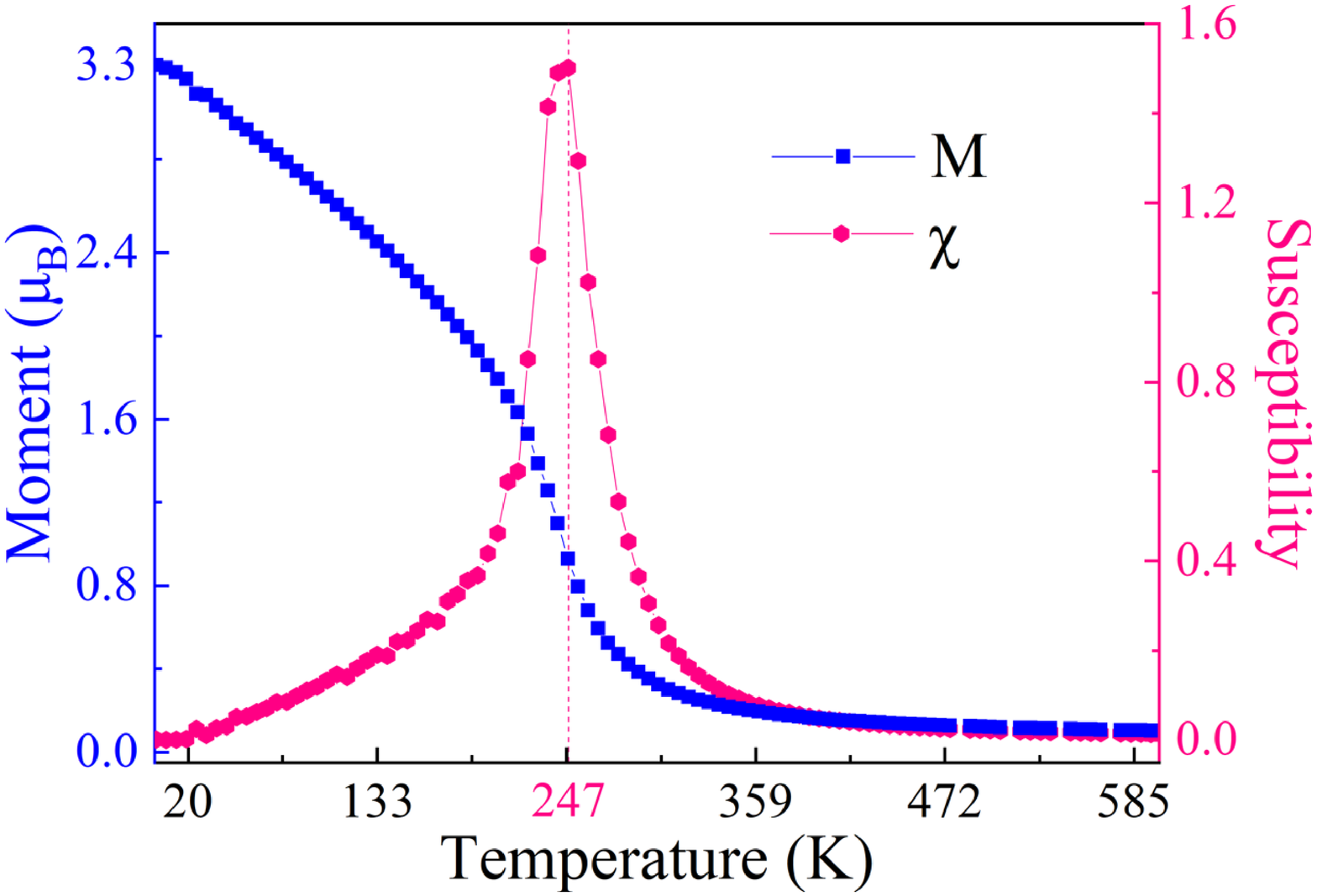}
\caption{Susceptibility $\chi$ and average magnetic moment $M$ as functions of temperature on a square-planar lattice for the MnN$_4$ monolayer.
 } \label{Tc}
\end{center}
\end{figure}

Ising model and Heisenberg model are usually used to describe the magnetic interactions of 2D magnetic materials.
Ising model is the limit of Heisenberg model with magnetic anisotropy going to infinity.
Because of the small magnetic anisotropy in real 2D ferromagnetic materials, Curie temperature is overestimated by Ising model.
So, the Heisenberg model is more precise, which has been successfully applied in CrI$_3$ and other synthesized ferromagnetic 2D materials \cite{Liu2016,Zhang2021,Guo2020}.
We adopt the Heisenberg model to evaluate the Curie temperature of MnN$_4$ monolayer.
The Hamiltonian is defined as
\begin{equation}
\label{Heisenberg}
H = J_{1}\sum_{<ij>}{\vec{S_i}} \cdot {\vec{S_j}} + J_{2}\sum_{\ll ij^{\prime} \gg}{\vec{S_i}} \cdot \vec{S_{j^{\prime}}} + A\sum_{i}(S_{iz})^2,
\end{equation}
in which the symbol $j$ and $j^\prime$ represent the neighboring sites of $i$ site in the square lattice, and $A$ is the single-site magnetic anisotropic energy.
To solve the Heisenberg Hamiltonian, the Monte Carlo method with a 60 $\times$ 60 $\times$ lattice is used.
The magnetization ($M$) and susceptibility ($\chi$ = $\frac{<\vec{M}^2> - <\vec{M}>^2}{k_BT}$) with respect to temperature are presented in Fig. \ref{Tc}. Both of them show a clear ferromagnetic phase transition at 247 K. So, the Curie temperatures ($T_{c}$) of MnN$_4$ monolayer is 247 K, which is a relatively high Curie temperature compared to the reported 2D ferromagnetic materials \cite{Guo2020}.

In summary, based on the first-principles calculations, we design a single-atom-thick manganese nitride MnN$_4$ with a porous square-planar network.
 Along Mn-N=N-Mn chain in the MnN$_4$ monolayer, the $d_{xz}$, $p_z$, $p_z$, and $d_{xz}$ orbitals are hybridized together to form the large $\pi$ orbitals. The $\pi$-d conjugation is the remarkable feature of the predicted MnN$_4$ monolayer, which is the origins of both structural stability and ferromagnetic couplings between two Mn atoms.
The robust ferromagnetism with the Curie temperature of 247 K is demonstrated by solving the Heisenberg Hamiltonian with the Monte Carlo method.
Therefore, our results not only predict a new high-temperature ferromagnetic 2D material but also propose a new mechanism of ferromagnetic exchange coupling.

This work was supported by the National R\&D Program of China (Grants No. 2016YFA0300503 and 2017YFA0302900), the National Natural Science Foundation of China (Grants Nos. 12274458, 11774420, 11974194), and the Research Funds of Renmin University of China (Grant No. 20XNLG19)

\bibliography{Ref}

\begin{thebibliography}{27}%
\makeatletter
\providecommand \@ifxundefined [1]{%
 \@ifx{#1\undefined}
}%
\providecommand \@ifnum [1]{%
 \ifnum #1\expandafter \@firstoftwo
 \else \expandafter \@secondoftwo
 \fi
}%
\providecommand \@ifx [1]{%
 \ifx #1\expandafter \@firstoftwo
 \else \expandafter \@secondoftwo
 \fi
}%
\providecommand \natexlab [1]{#1}%
\providecommand \enquote  [1]{``#1''}%
\providecommand \bibnamefont  [1]{#1}%
\providecommand \bibfnamefont [1]{#1}%
\providecommand \citenamefont [1]{#1}%
\providecommand \href@noop [0]{\@secondoftwo}%
\providecommand \href [0]{\begingroup \@sanitize@url \@href}%
\providecommand \@href[1]{\@@startlink{#1}\@@href}%
\providecommand \@@href[1]{\endgroup#1\@@endlink}%
\providecommand \@sanitize@url [0]{\catcode `\\12\catcode `\$12\catcode
  `\&12\catcode `\#12\catcode `\^12\catcode `\_12\catcode `\%12\relax}%
\providecommand \@@startlink[1]{}%
\providecommand \@@endlink[0]{}%
\providecommand \url  [0]{\begingroup\@sanitize@url \@url }%
\providecommand \@url [1]{\endgroup\@href {#1}{\urlprefix }}%
\providecommand \urlprefix  [0]{URL }%
\providecommand \Eprint [0]{\href }%
\providecommand \doibase [0]{http://dx.doi.org/}%
\providecommand \selectlanguage [0]{\@gobble}%
\providecommand \bibinfo  [0]{\@secondoftwo}%
\providecommand \bibfield  [0]{\@secondoftwo}%
\providecommand \translation [1]{[#1]}%
\providecommand \BibitemOpen [0]{}%
\providecommand \bibitemStop [0]{}%
\providecommand \bibitemNoStop [0]{.\EOS\space}%
\providecommand \EOS [0]{\spacefactor3000\relax}%
\providecommand \BibitemShut  [1]{\csname bibitem#1\endcsname}%
\let\auto@bib@innerbib\@empty
\bibitem [{\citenamefont {Ashton}\ \emph {et~al.}(2017)\citenamefont {Ashton},
  \citenamefont {Paul}, \citenamefont {Sinnott},\ and\ \citenamefont
  {Hennig}}]{Ashton2017}%
  \BibitemOpen
  \bibfield  {author} {\bibinfo {author} {\bibfnamefont {M.}~\bibnamefont
  {Ashton}}, \bibinfo {author} {\bibfnamefont {J.}~\bibnamefont {Paul}},
  \bibinfo {author} {\bibfnamefont {S.~B.}\ \bibnamefont {Sinnott}}, \ and\
  \bibinfo {author} {\bibfnamefont {R.~G.}\ \bibnamefont {Hennig}},\ }\href
  {\doibase 10.1103/PhysRevLett.118.106101} {\bibfield  {journal} {\bibinfo
  {journal} {Physical Review Letters}\ }\textbf {\bibinfo {volume} {118}},\
  \bibinfo {pages} {106101} (\bibinfo {year} {2017})}\BibitemShut {NoStop}%
\bibitem [{\citenamefont {Song}\ \emph {et~al.}(2010)\citenamefont {Song},
  \citenamefont {Ci}, \citenamefont {Lu}, \citenamefont {Sorokin},
  \citenamefont {Jin}, \citenamefont {Ni}, \citenamefont {Kvashnin},
  \citenamefont {Kvashnin}, \citenamefont {Lou}, \citenamefont {Yakobson},\
  and\ \citenamefont {Ajayan}}]{Song2010}%
  \BibitemOpen
  \bibfield  {author} {\bibinfo {author} {\bibfnamefont {L.}~\bibnamefont
  {Song}}, \bibinfo {author} {\bibfnamefont {L.}~\bibnamefont {Ci}}, \bibinfo
  {author} {\bibfnamefont {H.}~\bibnamefont {Lu}}, \bibinfo {author}
  {\bibfnamefont {P.~B.}\ \bibnamefont {Sorokin}}, \bibinfo {author}
  {\bibfnamefont {C.}~\bibnamefont {Jin}}, \bibinfo {author} {\bibfnamefont
  {J.}~\bibnamefont {Ni}}, \bibinfo {author} {\bibfnamefont {A.~G.}\
  \bibnamefont {Kvashnin}}, \bibinfo {author} {\bibfnamefont {D.~G.}\
  \bibnamefont {Kvashnin}}, \bibinfo {author} {\bibfnamefont {J.}~\bibnamefont
  {Lou}}, \bibinfo {author} {\bibfnamefont {B.~I.}\ \bibnamefont {Yakobson}}, \
  and\ \bibinfo {author} {\bibfnamefont {P.~M.}\ \bibnamefont {Ajayan}},\
  }\href {\doibase 10.1021/nl1022139} {\bibfield  {journal} {\bibinfo
  {journal} {Nano Letters}\ }\textbf {\bibinfo {volume} {10}},\ \bibinfo
  {pages} {3209} (\bibinfo {year} {2010})}\BibitemShut {NoStop}%
\bibitem [{\citenamefont {Lalmi}\ \emph {et~al.}(2010)\citenamefont {Lalmi},
  \citenamefont {Oughaddou}, \citenamefont {Enriquez}, \citenamefont {Kara},
  \citenamefont {Vizzini}, \citenamefont {Ealet},\ and\ \citenamefont
  {Aufray}}]{Lalmi2010}%
  \BibitemOpen
  \bibfield  {author} {\bibinfo {author} {\bibfnamefont {B.}~\bibnamefont
  {Lalmi}}, \bibinfo {author} {\bibfnamefont {H.}~\bibnamefont {Oughaddou}},
  \bibinfo {author} {\bibfnamefont {H.}~\bibnamefont {Enriquez}}, \bibinfo
  {author} {\bibfnamefont {A.}~\bibnamefont {Kara}}, \bibinfo {author}
  {\bibfnamefont {S.}~\bibnamefont {Vizzini}}, \bibinfo {author} {\bibfnamefont
  {B.}~\bibnamefont {Ealet}}, \ and\ \bibinfo {author} {\bibfnamefont
  {B.}~\bibnamefont {Aufray}},\ }\href {\doibase 10.1063/1.3524215} {\bibfield
  {journal} {\bibinfo  {journal} {Applied Physics Letters}\ }\textbf {\bibinfo
  {volume} {97}},\ \bibinfo {pages} {223109} (\bibinfo {year}
  {2010})}\BibitemShut {NoStop}%
\bibitem [{\citenamefont {Coleman}\ \emph {et~al.}(2011)\citenamefont
  {Coleman}, \citenamefont {Lotya}, \citenamefont {O'Neill}, \citenamefont
  {Bergin}, \citenamefont {King}, \citenamefont {Khan}, \citenamefont {Young},
  \citenamefont {Gaucher}, \citenamefont {De}, \citenamefont {Smith},
  \citenamefont {Shvets}, \citenamefont {Arora}, \citenamefont {Stanton},
  \citenamefont {Kim}, \citenamefont {Lee}, \citenamefont {Kim}, \citenamefont
  {Duesberg}, \citenamefont {Hallam}, \citenamefont {Boland}, \citenamefont
  {Wang}, \citenamefont {Donegan}, \citenamefont {Grunlan}, \citenamefont
  {Moriarty}, \citenamefont {Shmeliov}, \citenamefont {Nicholls}, \citenamefont
  {Perkins}, \citenamefont {Grieveson}, \citenamefont {Theuwissen},
  \citenamefont {McComb}, \citenamefont {Nellist},\ and\ \citenamefont
  {Nicolosi}}]{Coleman2011}%
  \BibitemOpen
  \bibfield  {author} {\bibinfo {author} {\bibfnamefont {J.~N.}\ \bibnamefont
  {Coleman}}, \bibinfo {author} {\bibfnamefont {M.}~\bibnamefont {Lotya}},
  \bibinfo {author} {\bibfnamefont {A.}~\bibnamefont {O'Neill}}, \bibinfo
  {author} {\bibfnamefont {S.~D.}\ \bibnamefont {Bergin}}, \bibinfo {author}
  {\bibfnamefont {P.~J.}\ \bibnamefont {King}}, \bibinfo {author}
  {\bibfnamefont {U.}~\bibnamefont {Khan}}, \bibinfo {author} {\bibfnamefont
  {K.}~\bibnamefont {Young}}, \bibinfo {author} {\bibfnamefont
  {A.}~\bibnamefont {Gaucher}}, \bibinfo {author} {\bibfnamefont
  {S.}~\bibnamefont {De}}, \bibinfo {author} {\bibfnamefont {R.~J.}\
  \bibnamefont {Smith}}, \bibinfo {author} {\bibfnamefont {I.~V.}\ \bibnamefont
  {Shvets}}, \bibinfo {author} {\bibfnamefont {S.~K.}\ \bibnamefont {Arora}},
  \bibinfo {author} {\bibfnamefont {G.}~\bibnamefont {Stanton}}, \bibinfo
  {author} {\bibfnamefont {H.-Y.}\ \bibnamefont {Kim}}, \bibinfo {author}
  {\bibfnamefont {K.}~\bibnamefont {Lee}}, \bibinfo {author} {\bibfnamefont
  {G.~T.}\ \bibnamefont {Kim}}, \bibinfo {author} {\bibfnamefont {G.~S.}\
  \bibnamefont {Duesberg}}, \bibinfo {author} {\bibfnamefont {T.}~\bibnamefont
  {Hallam}}, \bibinfo {author} {\bibfnamefont {J.~J.}\ \bibnamefont {Boland}},
  \bibinfo {author} {\bibfnamefont {J.~J.}\ \bibnamefont {Wang}}, \bibinfo
  {author} {\bibfnamefont {J.~F.}\ \bibnamefont {Donegan}}, \bibinfo {author}
  {\bibfnamefont {J.~C.}\ \bibnamefont {Grunlan}}, \bibinfo {author}
  {\bibfnamefont {G.}~\bibnamefont {Moriarty}}, \bibinfo {author}
  {\bibfnamefont {A.}~\bibnamefont {Shmeliov}}, \bibinfo {author}
  {\bibfnamefont {R.~J.}\ \bibnamefont {Nicholls}}, \bibinfo {author}
  {\bibfnamefont {J.~M.}\ \bibnamefont {Perkins}}, \bibinfo {author}
  {\bibfnamefont {E.~M.}\ \bibnamefont {Grieveson}}, \bibinfo {author}
  {\bibfnamefont {K.}~\bibnamefont {Theuwissen}}, \bibinfo {author}
  {\bibfnamefont {D.~W.}\ \bibnamefont {McComb}}, \bibinfo {author}
  {\bibfnamefont {P.~D.}\ \bibnamefont {Nellist}}, \ and\ \bibinfo {author}
  {\bibfnamefont {V.}~\bibnamefont {Nicolosi}},\ }\href {\doibase
  10.1126/science.1194975} {\bibfield  {journal} {\bibinfo  {journal}
  {Science}\ }\textbf {\bibinfo {volume} {331}},\ \bibinfo {pages} {568}
  (\bibinfo {year} {2011})}\BibitemShut {NoStop}%
\bibitem [{\citenamefont {Naguib}\ \emph {et~al.}(2011)\citenamefont {Naguib},
  \citenamefont {Kurtoglu}, \citenamefont {Presser}, \citenamefont {Lu},
  \citenamefont {Niu}, \citenamefont {Heon}, \citenamefont {Hultman},
  \citenamefont {Gogotsi},\ and\ \citenamefont {Barsoum}}]{Naguib2011}%
  \BibitemOpen
  \bibfield  {author} {\bibinfo {author} {\bibfnamefont {M.}~\bibnamefont
  {Naguib}}, \bibinfo {author} {\bibfnamefont {M.}~\bibnamefont {Kurtoglu}},
  \bibinfo {author} {\bibfnamefont {V.}~\bibnamefont {Presser}}, \bibinfo
  {author} {\bibfnamefont {J.}~\bibnamefont {Lu}}, \bibinfo {author}
  {\bibfnamefont {J.}~\bibnamefont {Niu}}, \bibinfo {author} {\bibfnamefont
  {M.}~\bibnamefont {Heon}}, \bibinfo {author} {\bibfnamefont {L.}~\bibnamefont
  {Hultman}}, \bibinfo {author} {\bibfnamefont {Y.}~\bibnamefont {Gogotsi}}, \
  and\ \bibinfo {author} {\bibfnamefont {M.~W.}\ \bibnamefont {Barsoum}},\
  }\href {\doibase 10.1002/adma.201102306} {\bibfield  {journal} {\bibinfo
  {journal} {Advanced Materials}\ }\textbf {\bibinfo {volume} {23}},\ \bibinfo
  {pages} {4248} (\bibinfo {year} {2011})}\BibitemShut {NoStop}%
\bibitem [{\citenamefont {Huang}\ \emph {et~al.}(2017)\citenamefont {Huang},
  \citenamefont {Clark}, \citenamefont {Navarro-Moratalla}, \citenamefont
  {Klein}, \citenamefont {Cheng}, \citenamefont {Seyler}, \citenamefont
  {Zhong}, \citenamefont {Schmidgall}, \citenamefont {McGuire}, \citenamefont
  {Cobden}, \citenamefont {Yao}, \citenamefont {Xiao}, \citenamefont
  {Jarillo-Herrero},\ and\ \citenamefont {Xu}}]{Huang2017}%
  \BibitemOpen
  \bibfield  {author} {\bibinfo {author} {\bibfnamefont {B.}~\bibnamefont
  {Huang}}, \bibinfo {author} {\bibfnamefont {G.}~\bibnamefont {Clark}},
  \bibinfo {author} {\bibfnamefont {E.}~\bibnamefont {Navarro-Moratalla}},
  \bibinfo {author} {\bibfnamefont {D.~R.}\ \bibnamefont {Klein}}, \bibinfo
  {author} {\bibfnamefont {R.}~\bibnamefont {Cheng}}, \bibinfo {author}
  {\bibfnamefont {K.~L.}\ \bibnamefont {Seyler}}, \bibinfo {author}
  {\bibfnamefont {D.}~\bibnamefont {Zhong}}, \bibinfo {author} {\bibfnamefont
  {E.}~\bibnamefont {Schmidgall}}, \bibinfo {author} {\bibfnamefont {M.~A.}\
  \bibnamefont {McGuire}}, \bibinfo {author} {\bibfnamefont {D.~H.}\
  \bibnamefont {Cobden}}, \bibinfo {author} {\bibfnamefont {W.}~\bibnamefont
  {Yao}}, \bibinfo {author} {\bibfnamefont {D.}~\bibnamefont {Xiao}}, \bibinfo
  {author} {\bibfnamefont {P.}~\bibnamefont {Jarillo-Herrero}}, \ and\ \bibinfo
  {author} {\bibfnamefont {X.}~\bibnamefont {Xu}},\ }\href {\doibase
  10.1038/nature22391} {\bibfield  {journal} {\bibinfo  {journal} {Nature}\
  }\textbf {\bibinfo {volume} {546}},\ \bibinfo {pages} {270} (\bibinfo {year}
  {2017})}\BibitemShut {NoStop}%
\bibitem [{\citenamefont {Gong}\ \emph {et~al.}(2017)\citenamefont {Gong},
  \citenamefont {Li}, \citenamefont {Li}, \citenamefont {Ji}, \citenamefont
  {Stern}, \citenamefont {Xia}, \citenamefont {Cao}, \citenamefont {Bao},
  \citenamefont {Wang}, \citenamefont {Wang}, \citenamefont {Qiu},
  \citenamefont {Cava}, \citenamefont {Louie}, \citenamefont {Xia},\ and\
  \citenamefont {Zhang}}]{Gong2017}%
  \BibitemOpen
  \bibfield  {author} {\bibinfo {author} {\bibfnamefont {C.}~\bibnamefont
  {Gong}}, \bibinfo {author} {\bibfnamefont {L.}~\bibnamefont {Li}}, \bibinfo
  {author} {\bibfnamefont {Z.}~\bibnamefont {Li}}, \bibinfo {author}
  {\bibfnamefont {H.}~\bibnamefont {Ji}}, \bibinfo {author} {\bibfnamefont
  {A.}~\bibnamefont {Stern}}, \bibinfo {author} {\bibfnamefont
  {Y.}~\bibnamefont {Xia}}, \bibinfo {author} {\bibfnamefont {T.}~\bibnamefont
  {Cao}}, \bibinfo {author} {\bibfnamefont {W.}~\bibnamefont {Bao}}, \bibinfo
  {author} {\bibfnamefont {C.}~\bibnamefont {Wang}}, \bibinfo {author}
  {\bibfnamefont {Y.}~\bibnamefont {Wang}}, \bibinfo {author} {\bibfnamefont
  {Z.~Q.}\ \bibnamefont {Qiu}}, \bibinfo {author} {\bibfnamefont {R.~J.}\
  \bibnamefont {Cava}}, \bibinfo {author} {\bibfnamefont {S.~G.}\ \bibnamefont
  {Louie}}, \bibinfo {author} {\bibfnamefont {J.}~\bibnamefont {Xia}}, \ and\
  \bibinfo {author} {\bibfnamefont {X.}~\bibnamefont {Zhang}},\ }\href
  {\doibase 10.1038/nature22060} {\bibfield  {journal} {\bibinfo  {journal}
  {Nature}\ }\textbf {\bibinfo {volume} {546}},\ \bibinfo {pages} {265}
  (\bibinfo {year} {2017})}\BibitemShut {NoStop}%
\bibitem [{\citenamefont {Mannix}\ \emph {et~al.}(2015)\citenamefont {Mannix},
  \citenamefont {Zhou}, \citenamefont {Kiraly}, \citenamefont {Wood},
  \citenamefont {Alducin}, \citenamefont {Myers}, \citenamefont {Liu},
  \citenamefont {Fisher}, \citenamefont {Santiago}, \citenamefont {Guest},
  \citenamefont {Yacaman}, \citenamefont {Ponce}, \citenamefont {Oganov},
  \citenamefont {Hersam},\ and\ \citenamefont {Guisinger}}]{Mannix2015}%
  \BibitemOpen
  \bibfield  {author} {\bibinfo {author} {\bibfnamefont {A.~J.}\ \bibnamefont
  {Mannix}}, \bibinfo {author} {\bibfnamefont {X.-F.}\ \bibnamefont {Zhou}},
  \bibinfo {author} {\bibfnamefont {B.}~\bibnamefont {Kiraly}}, \bibinfo
  {author} {\bibfnamefont {J.~D.}\ \bibnamefont {Wood}}, \bibinfo {author}
  {\bibfnamefont {D.}~\bibnamefont {Alducin}}, \bibinfo {author} {\bibfnamefont
  {B.~D.}\ \bibnamefont {Myers}}, \bibinfo {author} {\bibfnamefont
  {X.}~\bibnamefont {Liu}}, \bibinfo {author} {\bibfnamefont {B.~L.}\
  \bibnamefont {Fisher}}, \bibinfo {author} {\bibfnamefont {U.}~\bibnamefont
  {Santiago}}, \bibinfo {author} {\bibfnamefont {J.~R.}\ \bibnamefont {Guest}},
  \bibinfo {author} {\bibfnamefont {M.~J.}\ \bibnamefont {Yacaman}}, \bibinfo
  {author} {\bibfnamefont {A.}~\bibnamefont {Ponce}}, \bibinfo {author}
  {\bibfnamefont {A.~R.}\ \bibnamefont {Oganov}}, \bibinfo {author}
  {\bibfnamefont {M.~C.}\ \bibnamefont {Hersam}}, \ and\ \bibinfo {author}
  {\bibfnamefont {N.~P.}\ \bibnamefont {Guisinger}},\ }\href {\doibase
  10.1126/science.aad1080} {\bibfield  {journal} {\bibinfo  {journal}
  {Science}\ }\textbf {\bibinfo {volume} {350}},\ \bibinfo {pages} {1513}
  (\bibinfo {year} {2015})}\BibitemShut {NoStop}%
\bibitem [{\citenamefont {Groenewolt}\ and\ \citenamefont
  {Antonietti}(2005)}]{Groenewolt2005}%
  \BibitemOpen
  \bibfield  {author} {\bibinfo {author} {\bibfnamefont {M.}~\bibnamefont
  {Groenewolt}}\ and\ \bibinfo {author} {\bibfnamefont {M.}~\bibnamefont
  {Antonietti}},\ }\href {\doibase 10.1002/adma.200401756} {\bibfield
  {journal} {\bibinfo  {journal} {Advanced Materials}\ }\textbf {\bibinfo
  {volume} {17}},\ \bibinfo {pages} {1789} (\bibinfo {year}
  {2005})}\BibitemShut {NoStop}%
\bibitem [{\citenamefont {Kan}\ \emph {et~al.}(2013)\citenamefont {Kan},
  \citenamefont {Zhou}, \citenamefont {Sun}, \citenamefont {Kawazoe},\ and\
  \citenamefont {Jena}}]{Kan2013}%
  \BibitemOpen
  \bibfield  {author} {\bibinfo {author} {\bibfnamefont {M.}~\bibnamefont
  {Kan}}, \bibinfo {author} {\bibfnamefont {J.}~\bibnamefont {Zhou}}, \bibinfo
  {author} {\bibfnamefont {Q.}~\bibnamefont {Sun}}, \bibinfo {author}
  {\bibfnamefont {Y.}~\bibnamefont {Kawazoe}}, \ and\ \bibinfo {author}
  {\bibfnamefont {P.}~\bibnamefont {Jena}},\ }\href {\doibase
  10.1021/jz4017848} {\bibfield  {journal} {\bibinfo  {journal} {Journal of
  Physical Chemistry Letters}\ }\textbf {\bibinfo {volume} {4}},\ \bibinfo
  {pages} {3382} (\bibinfo {year} {2013})}\BibitemShut {NoStop}%
\bibitem [{\citenamefont {Kan}\ \emph {et~al.}(2014)\citenamefont {Kan},
  \citenamefont {Adhikari},\ and\ \citenamefont {Sun}}]{Kan2014}%
  \BibitemOpen
  \bibfield  {author} {\bibinfo {author} {\bibfnamefont {M.}~\bibnamefont
  {Kan}}, \bibinfo {author} {\bibfnamefont {S.}~\bibnamefont {Adhikari}}, \
  and\ \bibinfo {author} {\bibfnamefont {Q.}~\bibnamefont {Sun}},\ }\href
  {\doibase 10.1039/c3cp55146f} {\bibfield  {journal} {\bibinfo  {journal}
  {Physical Chemistry Chemical Physics}\ }\textbf {\bibinfo {volume} {16}},\
  \bibinfo {pages} {4990} (\bibinfo {year} {2014})}\BibitemShut {NoStop}%
\bibitem [{\citenamefont {Zhao}\ \emph {et~al.}(2013)\citenamefont {Zhao},
  \citenamefont {Wang},\ and\ \citenamefont {Zhang}}]{Zhao2013}%
  \BibitemOpen
  \bibfield  {author} {\bibinfo {author} {\bibfnamefont {M.}~\bibnamefont
  {Zhao}}, \bibinfo {author} {\bibfnamefont {A.}~\bibnamefont {Wang}}, \ and\
  \bibinfo {author} {\bibfnamefont {X.}~\bibnamefont {Zhang}},\ }\href
  {\doibase 10.1039/c3nr03323f} {\bibfield  {journal} {\bibinfo  {journal}
  {Nanoscale}\ }\textbf {\bibinfo {volume} {5}},\ \bibinfo {pages} {10404}
  (\bibinfo {year} {2013})}\BibitemShut {NoStop}%
\bibitem [{\citenamefont {Zhou}\ and\ \citenamefont {Sun}(2011)}]{Zhou2011}%
  \BibitemOpen
  \bibfield  {author} {\bibinfo {author} {\bibfnamefont {J.}~\bibnamefont
  {Zhou}}\ and\ \bibinfo {author} {\bibfnamefont {Q.}~\bibnamefont {Sun}},\
  }\href {\doibase 10.1021/ja204990j} {\bibfield  {journal} {\bibinfo
  {journal} {Journal of the American Chemical Society}\ }\textbf {\bibinfo
  {volume} {133}},\ \bibinfo {pages} {15113} (\bibinfo {year}
  {2011})}\BibitemShut {NoStop}%
\bibitem [{\citenamefont {Xu}\ and\ \citenamefont {Zhu}(2018)}]{Xu2018}%
  \BibitemOpen
  \bibfield  {author} {\bibinfo {author} {\bibfnamefont {Z.}~\bibnamefont
  {Xu}}\ and\ \bibinfo {author} {\bibfnamefont {H.}~\bibnamefont {Zhu}},\
  }\href {\doibase 10.1021/acs.jpcc.8b02323} {\bibfield  {journal} {\bibinfo
  {journal} {Journal of Physical Chemistry C}\ }\textbf {\bibinfo {volume}
  {122}},\ \bibinfo {pages} {14918} (\bibinfo {year} {2018})}\BibitemShut
  {NoStop}%
\bibitem [{\citenamefont {Zhao}\ and\ \citenamefont {Wang}(2020)}]{Zhao2020}%
  \BibitemOpen
  \bibfield  {author} {\bibinfo {author} {\bibfnamefont {K.}~\bibnamefont
  {Zhao}}\ and\ \bibinfo {author} {\bibfnamefont {Q.}~\bibnamefont {Wang}},\
  }\href {\doibase 10.1016/j.apsusc.2019.144620} {\bibfield  {journal}
  {\bibinfo  {journal} {Applied Surface Science}\ }\textbf {\bibinfo {volume}
  {505}},\ \bibinfo {pages} {144620} (\bibinfo {year} {2020})}\BibitemShut
  {NoStop}%
\bibitem [{\citenamefont {Kresse}\ and\ \citenamefont
  {Hafner}(1993)}]{PhysRevB.47.558}%
  \BibitemOpen
  \bibfield  {author} {\bibinfo {author} {\bibfnamefont {G.}~\bibnamefont
  {Kresse}}\ and\ \bibinfo {author} {\bibfnamefont {J.}~\bibnamefont
  {Hafner}},\ }\href {\doibase 10.1103/PhysRevB.47.558} {\bibfield  {journal}
  {\bibinfo  {journal} {Phys. Rev. B}\ }\textbf {\bibinfo {volume} {47}},\
  \bibinfo {pages} {558} (\bibinfo {year} {1993})}\BibitemShut {NoStop}%
\bibitem [{\citenamefont {Kresse}\ and\ \citenamefont
  {Furthm{\"{u}}ller}(1996)}]{PhysRevB.54.11169}%
  \BibitemOpen
  \bibfield  {author} {\bibinfo {author} {\bibfnamefont {G.}~\bibnamefont
  {Kresse}}\ and\ \bibinfo {author} {\bibfnamefont {J.}~\bibnamefont
  {Furthm{\"{u}}ller}},\ }\href {\doibase 10.1103/PhysRevB.54.11169} {\bibfield
   {journal} {\bibinfo  {journal} {Physical Review B}\ }\textbf {\bibinfo
  {volume} {54}},\ \bibinfo {pages} {11169} (\bibinfo {year}
  {1996})}\BibitemShut {NoStop}%
\bibitem [{\citenamefont {Perdew}\ \emph {et~al.}(1996)\citenamefont {Perdew},
  \citenamefont {Burke},\ and\ \citenamefont
  {Ernzerhof}}]{PhysRevLett.77.3865}%
  \BibitemOpen
  \bibfield  {author} {\bibinfo {author} {\bibfnamefont {J.~P.}\ \bibnamefont
  {Perdew}}, \bibinfo {author} {\bibfnamefont {K.}~\bibnamefont {Burke}}, \
  and\ \bibinfo {author} {\bibfnamefont {M.}~\bibnamefont {Ernzerhof}},\ }\href
  {\doibase 10.1103/PhysRevLett.77.3865} {\bibfield  {journal} {\bibinfo
  {journal} {Physical Review Letters}\ }\textbf {\bibinfo {volume} {77}},\
  \bibinfo {pages} {3865} (\bibinfo {year} {1996})}\BibitemShut {NoStop}%
\bibitem [{\citenamefont {Bl{\"{o}}chl}(1994)}]{PhysRevB.50.17953}%
  \BibitemOpen
  \bibfield  {author} {\bibinfo {author} {\bibfnamefont {P.~E.}\ \bibnamefont
  {Bl{\"{o}}chl}},\ }\href {\doibase 10.1103/PhysRevB.50.17953} {\bibfield
  {journal} {\bibinfo  {journal} {Physical Review B}\ }\textbf {\bibinfo
  {volume} {50}},\ \bibinfo {pages} {17953} (\bibinfo {year}
  {1994})}\BibitemShut {NoStop}%
\bibitem [{\citenamefont {Cococcioni}\ and\ \citenamefont
  {de~Gironcoli}(2005)}]{Cococcioni2005}%
  \BibitemOpen
  \bibfield  {author} {\bibinfo {author} {\bibfnamefont {M.}~\bibnamefont
  {Cococcioni}}\ and\ \bibinfo {author} {\bibfnamefont {S.}~\bibnamefont
  {de~Gironcoli}},\ }\href {\doibase 10.1103/PhysRevB.71.035105} {\bibfield
  {journal} {\bibinfo  {journal} {Physical Review B}\ }\textbf {\bibinfo
  {volume} {71}},\ \bibinfo {pages} {035105} (\bibinfo {year}
  {2005})}\BibitemShut {NoStop}%
\bibitem [{\citenamefont {Togo}\ and\ \citenamefont {Tanaka}(2015)}]{Togo2015}%
  \BibitemOpen
  \bibfield  {author} {\bibinfo {author} {\bibfnamefont {A.}~\bibnamefont
  {Togo}}\ and\ \bibinfo {author} {\bibfnamefont {I.}~\bibnamefont {Tanaka}},\
  }\href {\doibase 10.1016/j.scriptamat.2015.07.021} {\bibfield  {journal}
  {\bibinfo  {journal} {Scripta Materialia}\ }\textbf {\bibinfo {volume}
  {108}},\ \bibinfo {pages} {1} (\bibinfo {year} {2015})}\BibitemShut {NoStop}%
\bibitem [{\citenamefont {Martyna}\ \emph {et~al.}(1992)\citenamefont
  {Martyna}, \citenamefont {Klein},\ and\ \citenamefont
  {Tuckerman}}]{Martyna1992}%
  \BibitemOpen
  \bibfield  {author} {\bibinfo {author} {\bibfnamefont {G.~J.}\ \bibnamefont
  {Martyna}}, \bibinfo {author} {\bibfnamefont {M.~L.}\ \bibnamefont {Klein}},
  \ and\ \bibinfo {author} {\bibfnamefont {M.}~\bibnamefont {Tuckerman}},\
  }\href {\doibase 10.1063/1.463940} {\bibfield  {journal} {\bibinfo  {journal}
  {The Journal of Chemical Physics}\ }\textbf {\bibinfo {volume} {97}},\
  \bibinfo {pages} {2635} (\bibinfo {year} {1992})}\BibitemShut {NoStop}%
\bibitem [{\citenamefont {Sauvage}\ \emph {et~al.}(1982)\citenamefont
  {Sauvage}, \citenamefont {{De Backer}},\ and\ \citenamefont
  {Stymne}}]{Sauvage1982}%
  \BibitemOpen
  \bibfield  {author} {\bibinfo {author} {\bibfnamefont {F.~X.}\ \bibnamefont
  {Sauvage}}, \bibinfo {author} {\bibfnamefont {M.~G.}\ \bibnamefont {{De
  Backer}}}, \ and\ \bibinfo {author} {\bibfnamefont {B.}~\bibnamefont
  {Stymne}},\ }\href {\doibase 10.1016/0584-8539(82)80071-8} {\bibfield
  {journal} {\bibinfo  {journal} {Spectrochimica Acta Part A: Molecular
  Spectroscopy}\ }\textbf {\bibinfo {volume} {38}},\ \bibinfo {pages} {803}
  (\bibinfo {year} {1982})}\BibitemShut {NoStop}%
\bibitem [{\citenamefont {Wang}\ \emph {et~al.}(2021)\citenamefont {Wang},
  \citenamefont {Dong},\ and\ \citenamefont {Feng}}]{Wang2021}%
  \BibitemOpen
  \bibfield  {author} {\bibinfo {author} {\bibfnamefont {M.}~\bibnamefont
  {Wang}}, \bibinfo {author} {\bibfnamefont {R.}~\bibnamefont {Dong}}, \ and\
  \bibinfo {author} {\bibfnamefont {X.}~\bibnamefont {Feng}},\ }\href {\doibase
  10.1039/d0cs01160f} {\bibfield  {journal} {\bibinfo  {journal} {Chemical
  Society Reviews}\ }\textbf {\bibinfo {volume} {50}},\ \bibinfo {pages} {2764}
  (\bibinfo {year} {2021})}\BibitemShut {NoStop}%
\bibitem [{\citenamefont {Liu}\ \emph {et~al.}(2016)\citenamefont {Liu},
  \citenamefont {Sun}, \citenamefont {Kawazoe},\ and\ \citenamefont
  {Jena}}]{Liu2016}%
  \BibitemOpen
  \bibfield  {author} {\bibinfo {author} {\bibfnamefont {J.}~\bibnamefont
  {Liu}}, \bibinfo {author} {\bibfnamefont {Q.}~\bibnamefont {Sun}}, \bibinfo
  {author} {\bibfnamefont {Y.}~\bibnamefont {Kawazoe}}, \ and\ \bibinfo
  {author} {\bibfnamefont {P.}~\bibnamefont {Jena}},\ }\href {\doibase
  10.1039/C5CP04835D} {\bibfield  {journal} {\bibinfo  {journal} {Physical
  Chemistry Chemical Physics}\ }\textbf {\bibinfo {volume} {18}},\ \bibinfo
  {pages} {8777} (\bibinfo {year} {2016})}\BibitemShut {NoStop}%
\bibitem [{\citenamefont {Zhang}\ \emph {et~al.}(2021)\citenamefont {Zhang},
  \citenamefont {Wang}, \citenamefont {Guo}, \citenamefont {Li},\ and\
  \citenamefont {Wang}}]{Zhang2021}%
  \BibitemOpen
  \bibfield  {author} {\bibinfo {author} {\bibfnamefont {Y.}~\bibnamefont
  {Zhang}}, \bibinfo {author} {\bibfnamefont {B.}~\bibnamefont {Wang}},
  \bibinfo {author} {\bibfnamefont {Y.}~\bibnamefont {Guo}}, \bibinfo {author}
  {\bibfnamefont {Q.}~\bibnamefont {Li}}, \ and\ \bibinfo {author}
  {\bibfnamefont {J.}~\bibnamefont {Wang}},\ }\href {\doibase
  10.1016/j.commatsci.2021.110638} {\bibfield  {journal} {\bibinfo  {journal}
  {Computational Materials Science}\ }\textbf {\bibinfo {volume} {197}},\
  \bibinfo {pages} {110638} (\bibinfo {year} {2021})}\BibitemShut {NoStop}%
\bibitem [{\citenamefont {Guo}\ \emph {et~al.}(2020)\citenamefont {Guo},
  \citenamefont {Wang}, \citenamefont {Zhang}, \citenamefont {Yuan},
  \citenamefont {Ma},\ and\ \citenamefont {Wang}}]{Guo2020}%
  \BibitemOpen
  \bibfield  {author} {\bibinfo {author} {\bibfnamefont {Y.}~\bibnamefont
  {Guo}}, \bibinfo {author} {\bibfnamefont {B.}~\bibnamefont {Wang}}, \bibinfo
  {author} {\bibfnamefont {X.}~\bibnamefont {Zhang}}, \bibinfo {author}
  {\bibfnamefont {S.}~\bibnamefont {Yuan}}, \bibinfo {author} {\bibfnamefont
  {L.}~\bibnamefont {Ma}}, \ and\ \bibinfo {author} {\bibfnamefont
  {J.}~\bibnamefont {Wang}},\ }\href {\doibase 10.1002/inf2.12096} {\bibfield
  {journal} {\bibinfo  {journal} {InfoMat}\ }\textbf {\bibinfo {volume} {2}},\
  \bibinfo {pages} {639} (\bibinfo {year} {2020})}\BibitemShut {NoStop}%
\end{thebibliography}%

\end{document}